\documentclass[english,showpacs,floatfix,12pt]{revtex4}
\usepackage[dvips]{graphicx}
\usepackage{longtable}
\usepackage{amsmath,amssymb}
\usepackage{dcolumn}
\usepackage{latexsym}
\usepackage{babel}
\usepackage[latin1]{inputenc}
\usepackage{color}
\usepackage[colorlinks]{hyperref}

\newcommand\figcaption{\def\@captype{figure}\caption}

\def\dl{\delta}
\def\La{\Lambda}

\def\Om{\Omega}

\def\th{\theta}

\def\S{{\cal S}}

\begin{document}
\title{\Large Dynamical Constraints on the Cosmological Parameters}
\author{Ying-Qiu Gu}
\email{yqgu@fudan.edu.cn} \affiliation{School of Mathematical
Science, Fudan University, Shanghai 200433, China} \pacs{98.80.Jk,
98.80.Bp, 98.80.Cq, 95.36.+x}
\date{10th August 2017}

\begin{abstract}

In cosmology,  the cosmic curvature $K$ and the cosmological
constant $\La$ are two important parameters, and the values  have
strong influence on the behavior of the universe. In the context of
normal cosmology, under the ordinary assumptions of positive
mass-energy and initial negative pressure, we find the initial
singularity of the universe is certainly absent and we have $K=1$.
This means total spatial structure of the universe should be a
3-dimensional sphere $S^3$. For the cyclic cosmological model, we
have $\Lambda\lesssim 10^{-24} {\rm ly}^{-2}$. Obviously, such
constraints would be helpful for the researches on the properties of
dark matter and dark energy in cosmology. \vskip 0.3cm \noindent
{\bf Keywords:} {\sl cosmic curvature, cosmological constant,
negative pressure, dark matter, dark energy}
\end{abstract}

\maketitle

\section{Introduction}
\setcounter{equation}{0}

In cosmology, we have two important constants to be determined. They
are the cosmic curvature $K$ and the cosmological constant
$\La$\cite{lcdm}. Some characteristic parameters of the universe
such as the age $\cal T$, the Hubble's constant $H_0$, the total
mass density $\Om_{\rm tot}$ and so on have been measured to high
accuracy\cite{1,2,3,4,dt10,dt13}. To determine the cosmic curvature
$K$, the usual method is to transform the Friedmann equation into an
algebraic equation $\Om_K\equiv K\dot a^{-2}=\Om_{\rm tot}-1$, then
the case $K=0, \pm 1$ can be judged by examining the empirical data
$\Om_{\rm tot} >1, =1$ or $<1$.  By the observational data, we have
$\Om_{K}=-0.0020\pm 0.0047$, which is quite near the flat case and
hardly to be determined. In fact, it is easy to calculate that we
always have $\Om_{\rm tot}\approx 1$ for a young universe no matter
what case of the spatial curvature is. So this  criterion is quite
ambiguous.

The cosmological constant $\La$ has a dramatic history. Since
Einstein introduced $\La$ in 1917 to get a static and closed
universe, whether $\La=0$ or not has been debated many
times\cite{Lam1}-\cite{Lam4}. The dark matter and dark energy
attract the attention of the scientists all over the world and
become the hottest topics, which challenge the traditional standard
models of particles and cosmology. The usual description of dark
matter and dark energy is using the equation of state $P=w\rho$
together with $w=w(a)$ or $w=w(z)$, and many specific models to fit
the observational data are studied\cite{Lam1}-\cite{Lam10}. However,
the problem is far beyond solved\cite{lcdm}.

Since the Friedmann equation is a dynamical equation, the constants
can be hardly determined by statically analysis. To examine the
behavior of a dynamical equation according to the static algebraic
equation is usually unreliable and ambiguous. As shown below, we
find that the parameters $(K,\La)$ actually can be constrained by
qualitatively analyzing the properties of Friedmann equation under
quite ordinary conditions such as positive energy and initial
negative pressure. These conditions are based on a lot of well
established and widely studied models \cite{Lam7,Lam8,Lam9,Lam10},
which are of high reliability. Under such conditions, we find that
the initial singularity actually cannot reach, and some definite
constraints on the parameters $(K,\La)$ can be derived. The results
would be helpful for the researches on dark matter and dark energy
as well as the other issues in cosmology. These results may be
somehow different from the conventional ones. If the derivation is
right in logic, then we can only check the effectiveness of the
assumptions.

Some similar discussions with concrete gravitating sources were once
performed by many authors\cite{stein1}-\cite{cycl}. In
\cite{stein1,stein2,stein3}, the nonlinear scalar filed is
discussed, and the cyclic universe is obtained. In \cite{barr1}, a
number of exact cyclic solutions with normal dust and radiation were
obtained, and the exact solution with a ghost field and
electromagnetic field was derived in \cite{barr2}. The quantized
nonlinear spinor model and the trajectories were calculated in
\cite{gu2}. The Friedmann equation for some well-known dark energy
models were translated as the dynamics of Hamiltonian system by
introducing a potential function $V(a)$, and the evolution
trajectories are analyzed in \cite{Lam8}.

\section{notations and equations}
\setcounter{equation}{0}

In average sense, the universe is highly isotropic and
homogeneous, and the metric is described by
Friedmann-Robertson-Walker(FRW) metric. The corresponding line
element is usually given by
\begin{equation}
ds^2=d\tau^2-a^2(\tau)\left(\frac {d
r^2}{1-Kr^2}+r^2(d\th^2+\sin^2\th d\phi^2)\right), \label{eq1}
\end{equation}
where $K=1,~ 0$ and $-1$ correspond to the closed, flat and open
universe respectively, and $\tau$ is the comoving time. However, in
this form, the solution $a(\tau)$ can not be expressed by elementary
functions, and is nonanalytic as $a\to 0$ (e. g. $a\propto
\tau^{\frac 1 2}$ or $a\propto \tau^{\frac 2 3}$). Secondly, the
Friedmann equation
\begin{equation}
\dot a^2=-K+\frac 1 3\La a^2 + \frac{8\pi G}{3}\rho_{\rm m}
a^2,~~\left(\dot a=\frac {d a}{d\tau}\right) \label{eq2}
\end{equation}
includes singular term $\rho_{\rm m} a^2\to\infty$ as $a\to 0$,
which increases difficulties in discussion\cite{Lam8}. This problem
can be avoided by using conformal coordinate system. In this case,
the line element becomes
\begin{equation}
ds^2=a(t)^2\left(d t^2-dr ^2-\S (r)^2(d\th^2+\sin^2\th
d\phi^2)\right), \label{eq3}
\end{equation}
where $dt=a^{-1} d\tau$ is the conformal time,
\begin{equation} \S =\left \{ \begin{array}{ll}
  \sin r  & {\rm if} \quad  K=1,\\
   r   & {\rm if} \quad  K=0,\\
  \sinh r  & {\rm if} \quad  K=-1.
\end{array} \right. \label{eq4}
\end{equation}
Then the Friedmann equation (\ref{eq2}) becomes
\begin{equation}
a'^2=-Ka^2+\frac 1 3\La a^4 + \frac{8\pi G}{3}\rho_{\rm m}
a^4,\label{eq5}
\end{equation}
where the prime stands for $\frac d {dt}$, and $\rho_{\rm m}$ is the
total mass-energy density of all gravitating sources except for the
geometrical components $K$ and $\La$, but including particles,
radiation, dark matter, dark energy and so on. The total mass-energy
density $\rho_{\rm m}$ satisfies the energy conservation law
\begin{equation}
\frac d{da}({\rho_{\rm m} a^3})=-3Pa^2,\label{eq6}
\end{equation}
where $\rho_{\rm m}$ and $P$ is expressed as parametric functions of
$a$, and $a$ acts as parameter.

In \cite{Lam7}, the authors ranked 10 top dark energy models and 10
modified Friedmann equations according to the Bayesian information
criteria. In this paper, we are not concerned for the effectiveness
of these models, but concerned for the common features of their
Friedmann equation. The equations generally take the following form
\begin{equation}
H^2 = H^2_0\left(-\Om_f (1 + z)^4+\Om_{m}(1 + z)^3+ \Om_{K}(1 +
z)^2+\Om_\La+\Om_X(z) \right),\label{eq9}
\end{equation}
where
\begin{equation} H=\frac {da}{ad\tau}=\frac {a'(t)}{a^2},\qquad z=\frac
{a(t_a)}{a(t)}-1, \label{hz}\end{equation} $H$ is the Hubble's
parameter, $H_0=70\pm 4{\rm km~s^{-1} Mpc^{-1}}$ is the present
value of $H$, $z$ is the redshift and $t_a$ the present time,
$(\Om_m, \Om_K, \Om_\La)$ are the dimensionless energy densities
corresponding to common matter, curvature and cosmological constant
respectively. $\Om_f>0$ is caused by potential of fields such as
spinors \cite{gu2} and the Casimir effect of massless
scalar\cite{Lam7}, $\Om_X(z)>0$ stands for the energy density of
other dark matter and dark energy as well as the effects of
modification of general relativity, which is different from the
previous terms. Since (\ref{eq9}) is actually a dynamical equation
which can not be analyzed as algebraic equation, substituting
(\ref{hz}) into (\ref{eq9}), we convert it into the dynamical from.
More generally, the corresponding Friedmann equation should take the
following form
\begin{equation}
a'^2 =F(a),\quad F\equiv -\rho_f+2 R a -K a^2 +\frac 1 3 \La
a^4+X(a),\label{eq11}
\end{equation}
where $\rho_f>0$ corresponds to negative potential of fields, $R$
corresponds to the total comoving mass-energy density including
mass-energy density $\rho_0$ of particles and dark matter, which is
a constant\cite{gu2}
\begin{equation}
R=\frac{4\pi} 3 \varrho_m,\qquad \varrho_m=\rho_0 a^3. \label{rrho}
\end{equation}
$X(a)$ corresponds to the unknown parts of the dark matter and dark
energy, which is different from the previous terms and usually take
small values. The following discussion shows, the concrete form of
$X(a)$ is not important, and only its asymptotic properties as $a\to
+0$ have influence on the results.

By (\ref{eq11}) we find $R$ is the mean scale of the universe, which
can be used as length unit. Comparing (\ref{eq11}) with (\ref{eq5}),
we get total mass-energy density in the usual sense
\begin{eqnarray} \rho_{\rm m} &=&\frac 3{8\pi G a^4}\left(F(a)+K a^2
-\frac 1 3 \La
a^4\right),\label{rho}\\
&=&\frac 3{8\pi G a^4}\left(-\rho_f+2 R a +X(a)\right). \label{eq12}
\end{eqnarray}
Substituting (\ref{eq12}) into (\ref{eq6}), we get the pressure
\begin{eqnarray} P=-\frac 1{8\pi G a^4}\left(\rho_f
+{X'(a)a-X(a)}\right). \label{eq13}\end{eqnarray} Obviously the
derivatives of pressure and potential correspond to ordinary forces
which should be finite, so $P$ should be at least continuous. Then
by (\ref{eq13}) and the continuity of pressure $P$, we have at least
$X(a)\in C^1$.

\section{assumptions and analysis}
\setcounter{equation}{0}

The following discussion is based on Friedamnn equation (\ref{eq11})
and two assumptions upon energy density $\rho_{\rm m}$ and pressure
$P$. By their physical meaning and the observational facts, we have
two conclusions:

{\bf A1.} The total mass-energy density is always positive, namely
\begin{equation}
\rho_{\rm m}>0,\quad (\forall a>0).\label{rhom}
\end{equation}

{\bf A2.} The pressure $P<0$ when the universe is small,
\begin{equation}
P<0,\quad (a\to +0). \label{pngt}
\end{equation}

(\ref{rhom}) is a result of positive definite Hamiltonian of matter
or positive energy of matter, and (\ref{pngt}) is an observational
fact. In cosmology, although we call $P$ pressure, but it is
actually a variable not only including the usual positive term $\sum
m_n u^k_n u^l_n\dl(\vec x-\vec X_n)\sqrt{1-v^2_n}$ corresponding to
thermal movement of micro particles, but also including the
potentials of all fields\cite{gu2, Lam7}. So $P<0$ is reasonable in
physics. For example, the nonlinear spinors\cite{eng,gu0} and most
famous  dark energy models\cite{Lam7} all include negative pressure.
(A1) and (A2) are the basic assumptions for the following
discussion.

In what follows we prove
\begin{equation}
F(a)<0,\quad (a\to +0). \label{fngt}
\end{equation}
In the case of $|X(0)|<\infty$, by the condition (A2) $P<0(a\to 0)$,
we have
\begin{eqnarray} P\to -\frac 1{8\pi G
a^4}\left(\rho_f-X(0)\right)<0.\label{eq14.0}\end{eqnarray}
Consequently, by the definition of $F(a)$ in (\ref{eq11}), we get
\begin{equation} F(0)=-\rho_f+ X(0)<0. \label{eq14}
\end{equation}
In the case of $X\to {X_0} a^{-n},(a\to+0,n>0)$, which corresponds
to some nonlinear potentials\cite{eng,gu0}, we have
\begin{eqnarray}
P \to \frac {(n+1)X_0}{8\pi G a^{4+n}},\qquad (a\to +0).
\end{eqnarray}
by $P<0$ we find $X_0<0$. According to the definition of $F(a)$ in
(\ref{eq11}), we get
\begin{equation} F\to \frac {X_0}{ a^n}<0, \qquad (a\to +0). \label{Deq14*}
\end{equation}
Then we prove (\ref{fngt}) holds in all cases.

(\ref{fngt}) implies an important conclusion: {\bf The evolution of
the universe can not reach the initial singularity}. Now we check
this result. For the practical solution of Friedmann equation, we
should have $F(a)=a'^2\ge 0$. By (\ref{fngt}) and the continuity of
$F(a)$, the equation $F(a)=0$ certainly have a positive root
\begin{equation} 0<a_0\approx \frac {\rho_f}{2R}\ll R. \label{root}
\end{equation}

If $F(a)=0$ only has this positive real root $a_0$, then $F(a)$ can
be expressed as
\begin{equation} F(a)=(a-a_0) A(a),\quad (A>0,~ \forall a\ge a_0). \label{eq15}
\end{equation}
If $F(a)=0$ has a series of different positive roots
$0<a_0<a_1<a_2<\cdots$, then $F(a)$ for the practical universe can
be expressed as
\begin{equation} F(a)=(a-a_0)(a_1-a) B(a),\quad (B>0,~a_0\le a\le a_1). \label{eq16}
\end{equation}
Since Friedmann equation is an equation in average sense, the
multiple roots are meaningless in physics.

\begin{figure}
\centering
\includegraphics[width=12cm]{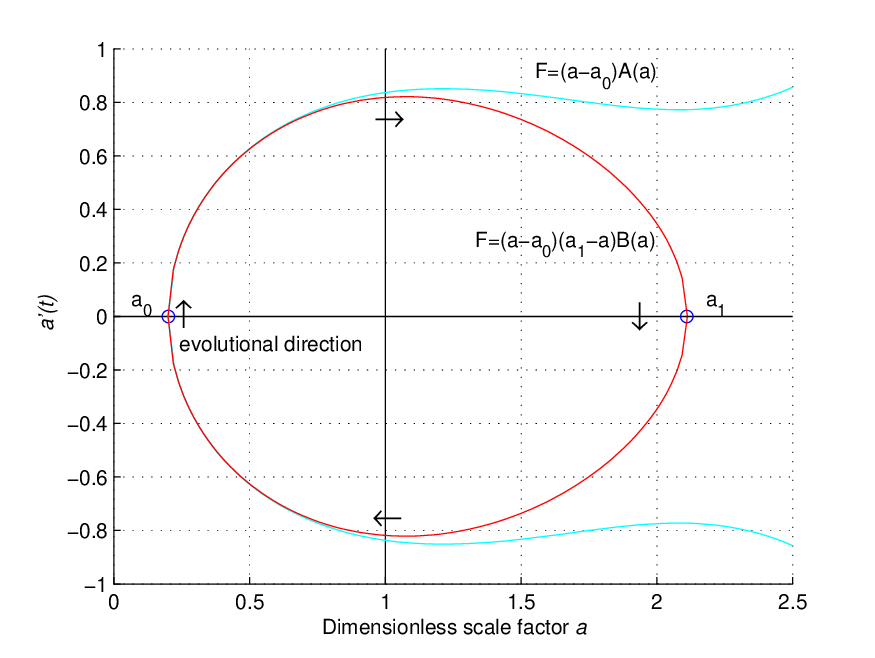}
\caption{The phase trajectories of Friedmann equation (\ref{eq11})}
\label{fig}
\end{figure}

The connected phase trajectories $a \sim a'$ of dynamical equation
(\ref{eq11}) with (\ref{eq15}) or (\ref{eq16}) are displayed in
FIG.\ref{fig}, in which we have set the mean scale $R=1$.
(\ref{eq16}) corresponds to the cyclic cosmological model, and
(\ref{eq15}) to the noncyclic one. We set the time origin $t=0$ at
the turning point $a(0)=a_0$. Form Fig.\ref{fig} we find the initial
singularity is absent, i.e. $a$ can not reaches $0$ point.

Substituting (\ref{eq15}) or (\ref{eq16}) into (\ref{rho}) and
letting $a=a_0$, we get a decisive criterion for $K$. In both cases,
we have
\begin{eqnarray} \rho_{\rm m}(a_0) = \frac 3{8\pi G
a^2_0}\left(K -\frac 1 3 \La a^2_0\right)>0. \label{eq19}
\end{eqnarray}
Since $\La\ge 0$ in cosmology, by (\ref{eq19}) we certainly have
$K=1$ due to $\rho_{\rm m}>0$. Then we get another conclusion: {\bf
The space of the universe should be a closed 3-dimensional sphere
$S^3$}.

In the cyclic closed case (\ref{eq16}), we have an estimation of
upper bound for the cosmological constant $\La$. Substituting
(\ref{eq16}) into (\ref{rho}) and letting $a=a_1$, we have
\begin{eqnarray} \rho_{\rm m}(a_1) = \frac 3{8\pi G
a^2_1}\left(1 -\frac 1 3 \La a^2_1\right)>0. \label{eq20}
\end{eqnarray}
Consequently, we get \begin{eqnarray} 0\le \La<\frac 3
{a_1^2}\approx \frac 3{4 R^2}\sim 10^{-24} {\rm ly}^{-2}.
\label{eq21}
\end{eqnarray}Then we get the third conclusion: {\bf For the cyclic
cosmological model, we have a tiny or vanishing $\La\lesssim
10^{-24} {\rm ly}^{-2}$.}

The noncyclic model with closed space and positive $\La$ can not be
ruled out by similar qualitative analysis. However such model might
be inconsistent with the isotropy and homogeneity of the present
universe, because the universe should be heavily anisotropy and
inhomogeneity before the turning point $t<0$ due to the lack of
initial causality among remote parts, and some information should be
kept today.

\section{discussions and conclusions}
\setcounter{equation}{0}

To sum up, by qualitatively analyzing the dynamical behavior of the
general Friedmann equation and the relations between parameters, we
can get some definite constraints on $(K,\La)$.  We find that only
the cyclic and closed cosmological model with a tiny or vanishing
$\La$ is natural and reasonable in physics. The other cases may
include nonphysical effects or contradictions. These constraints
would be helpful for the research of some issues of cosmology.
Unless the conditions (\ref{rhom}), (\ref{pngt}) or Friedmann
equation is violated seriously, the conclusions are certainly right.

\begin{acknowledgments}
The author is grateful to Prof. Ta-Tsien Li and Prof, Tie-Hu Qin for
their encouragement.
\end{acknowledgments}


\begin{thebibliography}{99}
\bibitem{lcdm} Ph. Bull1a, Y. Akrami,  {\em et al., Beyond $\Lambda$CDM: Problems, solutions, and the road
ahead}, Physics of the Dark Universe 12 (2016) 56-99,
arXiv:1512.05356v2
\bibitem{1} A. G. Riess {\em et al.} (Supernova Search Team), Astron. J. 116,
1009 (1998),  astro-ph/9805201.
\bibitem{2} S. Perlmutter {\em et al.} (Supernova Cosmology Project), Astrophys. J. 517, 565
(1999).
\bibitem{3} N. Spergel {\em et al.} (WMAP), Astrophys. J. Suppl. 148, 175 (2003),  astro-ph/0302209.
\bibitem{4} M. Tegmark {\em et al.} (SDSS), Phys. Rev. D69, 103501 (2004), astro-ph/0310723.
\bibitem{dt10} J. Dunkley {\em et al.} {\em The Atacama Cosmology
Telescope: Cosmological Parameters from the 2008 Power Spectra},
Astrophys. J. Vol.739, No.1, arXiv:1009.0866 [astro-ph.CO]
\bibitem{dt13} J. L. Sievers {\em et al.} {\em The Atacama Cosmology Telescope: Cosmological parameters from three seasons of
data},  J. Cosm. Astro. Phys., Vol.2013, Oct. 2013 arXiv:1301.0824v3
[astro-ph.CO]
\bibitem{Lam1} V. Sahni, {\em The cosmological constant problem and quintessence}, Class.Quant.Grav. 19(2002) 3435-3448, astro-ph/0202076
\bibitem{prm} J. Fonseca, R. Maartens, M. G. Santos, {\em Probing the primordial
Universe with MeerKAT and DES}, Mon. Not. Roy. Astron. Soc., 466(3),
2780 (2017). arXiv:1611.01322v2
\bibitem{Lam2} S. E. Deustua, {\em et al, Cosmological Parameters, Dark Energy and Large Scale Structure},\\ astro-ph/0207293
\bibitem{Lam3} P. J. E. Peebles, B. Ratra, {\em The Cosmological Constant and Dark Energy}, Rev. Mod. Phys. 75(2003) 559-606, astro-ph/0207347
\bibitem{Lam4} M. S. Turner, D. Huterer, {\em Cosmic Acceleration, Dark Energy and Fundamental Physics}, J.Phys.Soc.Jap.76:111015,2007, arXiv:0706.2186
\bibitem{Lam5} T. Padmanabhan, {\em Dark Energy: the Cosmological Challenge of the Millennium}, Curr. Sci. 88(2005) 1057, astro-ph/0411044
\bibitem{Lam6} M. Ishak, {\em Remarks on the formulation of the cosmological constant/dark energy problems}, Found.Phys.37:1470-1498,2007, astro-ph/0504416
\bibitem{Lam7} M. Szydlowski, A. Kurek, A. Krawiec, {\em Top ten accelerating cosmological models}, Phys. Lett. B642(2006) 171-178, astro-ph/0604327
\bibitem{Lam8} Marek Szydlowski, {\em Cosmological zoo - accelerating models with dark energy},\\ JCAP0709:007,2007, astro-ph/0610250
\bibitem{Lam9} E. J. Copeland, M. Sami,  S. Tsujikawa, {\em Dynamics of dark energy}, IJMPD {15},
1753-1936(2006), hep-th/0603057.
\bibitem{Lam10} E. V. Linder, {\em Theory Challenges of the Accelerating Universe}, J.Phys.A40:6697,2007, astro-ph/0610173
\bibitem{stein1} P. J. Steinhardt, N.Turok, {\em A Cyclic Model of the
Universe}, Science 296, 1436(2002), arXiv:hep-th/0111030
\bibitem{stein2} P. J. Steinhardt, N.Turok, {\em The Cyclic Model
Simplified}, New Astron.Rev. 49 (2005)43-57, astro-ph/0404480
\bibitem{stein3} J. Khoury, P. J. Steinhardt, N.Turok, {\em
Designing Cyclic Universe Models}, Phys. Rev. Lett.
92:031302(2004), hep-th/0307132
\bibitem{barr1} J. D. Barrow, M. P. Dabrowski, {\em Oscillating universes}, Mon.
Not. Astron. Soc. 275, 850-862(1995).
\bibitem{barr2}  J. D. Barrow, D. Kimberly, J. Magueijo, {\em Bouncing Universes with
Varying Constants}, arXiv:astro-ph/0406369v1
\bibitem{gu2} Y. Q. Gu, {\em A Cosmological Model with Dark Spinor Source},
Int. J. Mod. Phys. A22:4667-4678(2007), gr-qc/0610147.
\bibitem{cycl} P. H. Frampton, {\em ON CYCLIC UNIVERSES}, arXiv:astro-ph/0612243v1
\bibitem{eng} Y. Q. Gu, {\em The Vierbein Formalism and Energy-Momentum Tensor of Spinors}, arXiv:gr-qc/0612106
\bibitem{gu0} Y. Q. Gu, {\em Nonlinear Spinors as the Candidate of Dark Matter},
arXiv:0806.4649v2

\end{thebibliography}
\end{document}